\documentclass[
    twocolumn,
	prd,
	amssymb,
	preprintnumbers,
	secnumarabic,
	nofootinbib,
	superscriptaddress]{revtex4-1}

\pdfoutput=1

\usepackage{graphicx}
\usepackage{enumitem}
\usepackage{latexsym}
\usepackage{amsfonts}
\usepackage{amssymb}
\usepackage{color}
\usepackage{amsmath}
\usepackage{slashed}
\usepackage{dcolumn}
\usepackage{verbatim}
\usepackage{float}
\usepackage{multirow}
\usepackage{xspace}
\usepackage[normalem]{ulem}
\usepackage{hyperref}

\definecolor{hypershade}{rgb}{0.8,0.3,0.3}
\hypersetup{
  pdfauthor={Nirmal Raj},
  pdftitle={Cosmological and astrophysical probes of dark baryons},
  pdfsubject={Cosmological and astrophysical probes of dark baryons},
  colorlinks=true,
  citecolor=magenta,
  urlcolor=blue,
  linkcolor=hypershade
}

\newcommand{\gsim}{\gtrsim}
\newcommand{\lsim}{\lesssim}
\newcommand{\ra}{\rightarrow}

\def\Oc{\mathcal{O}}

\newcommand{\acro}[1]{\textsc{\MakeLowercase{#1}}} 

\renewcommand{\tilde}{\widetilde} 

\newcommand{\beq}{\begin{equation}}
\newcommand{\eeq}{\end{equation}}
\newcommand{\bea}{\begin{eqnarray}}
\newcommand{\eea}{\end{eqnarray}}
\newcommand{\nn}{\nonumber}

\def\mdm{m_\chi}

\allowdisplaybreaks

\begin{document}

\title{Cosmological and astrophysical probes of dark baryons}

\author{David McKeen}
\email{mckeen@triumf.ca}
\affiliation{TRIUMF, 4004 Wesbrook Mall, Vancouver, BC V6T 2A3, Canada}

\author{Maxim Pospelov}
\email{pospelov@umn.edu}
\affiliation{School of Physics and Astronomy, University of Minnesota, Minneapolis, MN 55455, USA}
\affiliation{William I. Fine Theoretical Physics Institute, School of Physics and Astronomy, University of Minnesota, Minneapolis, MN 55455, USA}

\author{Nirmal Raj}
\email{nraj@triumf.ca}
\affiliation{TRIUMF, 4004 Wesbrook Mall, Vancouver, BC V6T 2A3, Canada}

\date{\today}

\begin{abstract}
We examine the cosmological and astrophysical signatures of a ``dark baryon,'' a neutral fermion that mixes with the neutron. 
As the mixing is through a higher-dimensional operator at the quark level, production of the dark baryon at high energies is enhanced so that its abundance in the early universe may be significant. 
Treating its initial abundance as a free parameter, we derive new, powerful limits on the properties of the dark baryon. 
Primordial nucleosynthesis and the cosmic microwave background provide strong constraints due to the inter-conversion of neutrons to dark baryons through their induced transition dipole, and due to late decays of the dark baryon. 
Additionally, neutrons in a neutron star could decay slowly to dark baryons, providing a novel source of heat that is constrained by measurements of pulsar temperatures. 
Taking all the constraints into account, we identify parameter space where the dark baryon can be a viable dark matter candidate and discuss promising avenues for probing it.
\end{abstract}

\maketitle

\section{Introduction}
\label{sec:intro}

New states with masses around a GeV that mix with the standard model (SM) baryons have been considered recently for a number of compelling reasons, such as the baryon asymmetry of the universe~\cite{McKeen:2015cuz,Aitken:2017wie,Babu:2013yca,*Allahverdi:2017edd,*Elor:2018twp,*Nelson:2019fln,*Alonso-Alvarez:2019fym,Bringmann:2018sbs}, models of dark matter~\cite{McKeen:2015cuz,Karananas:2018goc,ClineCornellCosmo,Fornal:2020poq}, mirror matter scenarios~\cite{Berezhiani:2005hv,*Berezhiani:2015afa,*Berezhiani:2018eds}, the neutron lifetime anomaly~\cite{FornalGrinstein1801,FornalGrinsteinReview,BerezhianinHDK,ClineCornellCosmo,ElahiNonAbelCosmo}, the recent XENON1T excess~\cite{X1TexcessHportal}, $21~{\rm cm}$ cosmology~\cite{Johns:2020mmo,*Johns:2020rtp}, and general baryon-number violating phenomenology~\cite{Heeck:2020nbq,Fajfer:2020tqf}.
In this paper we investigate the cosmological and astrophysical effects of a simple, minimal version: a single ``dark baryon'' $\chi$ that carries baryon number $B=1$ and mixes with the neutron.

The phenomenology of our setup is characterized by just two parameters: the $\chi$ mass, $m_\chi$, and the $n$-$\chi$ mixing angle, $\theta$. This simple possibility has been singled out as a potential solution of the $4\sigma$ discrepancy between ``bottle'' and ``beam'' measurements of the neutron lifetime (although potentially introducing tension with determinations of the nucleon axial vector coupling~\cite{Marciano_gA}). Furthermore, such a new state could have a long lifetime---either absolutely stable if $m_\chi<m_p+m_e$ or cosmologically long-lived if $m_\chi>m_p+m_e$ and $\theta\ll 1$---and is therefore a potential dark matter (DM) candidate.

The simplicity of the ``minimal" dark baryon model at low energy is somewhat deceptive: processes in neutron stars (NS) tend to equilibrate $n$ and $\chi$ \cite{McKeenNelsonReddyZhouNS,SheltonNS,*MottaNS,*EllisPattavinaNS,*Goldman:2013qla}, which together with the observations of the most massive NSs seems to require additional repulsive interaction between $\chi$ particles \cite{McKeenNelsonReddyZhouNS}. This may in turn require, e.g. a composite nature of $\chi$ itself and/or the existence of a new ``dark vector" force. 
Both these scenarios hint at the existence of a more complicated dark sector (DS), with some states being in the sub-GeV range.
Another important consequence of the compositeness of normal hadrons is the dimensionality of interaction that induces the mixing angle $\theta$. 
The lowest dimension of the operator that seems to mediate such a transition is six, e.g. symbolically $\bar\chi(qqq)$, which in turn implies the enhancement of SM-DS inter-conversion at high temperatures. 
If the cutoff scale of this operator is not too far from the weak scale, a complete early thermalization of the SM and DS is plausible, with subsequent chemical decoupling of the two sectors. 
This motivates us to consider possible cosmological manifestations of the dark baryon model, in addition to one that was already pointed out in the literature, namely that $\chi$ might be a good candidate for a DM particle~\cite{McKeen:2015cuz,Karananas:2018goc,ClineCornellCosmo}. 
Other cosmological effects were further studied in Refs.~\cite{Karananas:2018goc,ClineCornellCosmo,ElahiNonAbelCosmo}. 

Here we examine key cosmological consequences of $\chi$ treating its initial abundance as a free parameter in a wide range: $n_\chi \sim O(0.01-5)\times n_{\rm baryon}$.
We will see that $\chi$ can affect big bang nucleosynthesis (BBN) of light nuclei by altering neutron-proton freeze-out, providing a source of neutrons after the deuterium bottleneck, or by decaying and breaking up light nuclei electromagnetically. 
In addition, $\chi$ that decays with a lifetime between about $10^{7}-10^{26}~{\rm s}$ can disrupt the excellent agreement between standard predictions and observations of the cosmic microwave background (CMB) frequency spectrum and temperature anisotropies.
Further, we identify a novel late-time heating mechanism of NSs: if the decay of stellar neutrons into dark baryons is very slow, equilibrium between the two species is not achieved over the NS lifetime, and heat is generated by the removal of neutrons from their Fermi sea followed by their replacement.
We exploit this phenomenon to place powerful constraints using NS temperature measurements.

Existing limits on this model in the $\theta\gtrsim 10^{-10}$ region come from a number of sources. The stability of the proton and $^9{\rm Be}$ require $m_\chi>938.0~\rm MeV$~\cite{McKeen:2015cuz,ExoticNucleiDecay}. The exotic neutron decay modes $n\to\chi\gamma$~\cite{ucnGammaLANL} and $n\to\chi e^+e^-$\cite{ucnaEE} have also been searched for, with null results providing limits on the parameter space. The low energy Borexino spectrum has also been recast as a search for hydrogen decay which can occur in this model when $m_\chi<m_p+m_e$~\cite{HDK}. As already mentioned, there is strong sensitivity to this model for $m_\chi\lesssim 1.2~\rm GeV$ that come from the existence of heavy NSs~\cite{McKeenNelsonReddyZhouNS,SheltonNS,*MottaNS,*EllisPattavinaNS,*Goldman:2013qla}.
Other signatures include the diffuse $\gamma$-ray background~\cite{BerezhianinHDK} and annihilation with nucleons in the case that the dark matter is comprised of $\bar\chi$ with $B=-1$~\cite{Davoudiasl:2010am,*Davoudiasl:2011fj,*GaoJinAnnihil,*MarfatiaAnnihil}.

This paper is laid out as follows.
In Section~\ref{sec:model} we introduce the dark baryon model arising through mass mixing with the neutron, and provide expressions for relevant rates.
In Section~\ref{sec:signals} we outline the effects of $\chi$ on BBN abundances and CMB measurements.
We treat $\chi$ both lighter and heavier than neutrons: the first region contains parameter space that explains the neutron lifetime puzzle, and the second region is so far unexplored in the presence of self-interactions obviating NS mass limits. 
In this section we also describe heating of NSs via cosmologically slow decays of constituent neutrons into $\chi$.
In Section~\ref{sec:limits} we derive constraints for initial abundances of $\chi$ equal to the DM abundance and 1\% of the baryon abundance, which lead to different phenomenologies.
In Section~\ref{sec:concs} we provide discussion on future prospects from cosmological, astrophysical, and terrestrial probes.

\section{Dark baryon model}
\label{sec:model}

The model consists of a charge-neutral fermion $\chi$ with $B = 1$ which mixes with the neutron,
\beq
\mathcal{L} \supset -\delta (\bar\chi n + \bar n \chi)~. 
\label{eq:Lag}
\eeq
In principle, a $\bar\chi i\gamma_5 n$ mixing term could exist as well. 
It would modify some of the phenomenological consequences discussed below, but not dramatically. 
We assume, effectively, that $\chi$ has the same internal parity as the neutron, and that interaction (\ref{eq:Lag}) is parity-symmetric.

This effective hadron-level Lagrangian (\ref{eq:Lag}) could arise from a number of quark-level UV completions; a simple one involves a scalar diquark with $ud$ and $d\chi$ couplings (see, e.g.~\cite{Arnold:2012sd,McKeen:2015cuz}). For $\delta \ll |\Delta m|$, where $\Delta m = m_n - m_\chi$, the $n$-$\chi$ mixing angle is $\theta = \delta/\Delta m$.

Diagonalization of the mass matrix results in a transition magnetic dipole,
\beq
\mathcal{L}_{\rm eff} \supset \frac{\mu_n}{2} \theta \bar\chi \sigma^{\mu \nu} n F_{\mu \nu}+{\rm H.c.}
\label{eq:tmm}
\eeq
In this expression, $\mu_n=-1.91\mu_N$ is the neutron magnetic dipole moment, with $\mu_N=e/(2m_n)\simeq 0.1~e\,{\rm fm}$ the nuclear magneton. For $\mdm < m_n$, the decay $n \to \chi \gamma$ occurs via \eqref{eq:tmm} with the rate
\beq
\Gamma_{n\ra \chi \gamma} = \theta^2 \left(\frac{\mu_n}{\mu_N}\right)^2\frac{\alpha\omega^3}{m_n^2}~,
\label{eq:nchigamma}
\eeq
where $\omega=(m_n/2)(1-m_\chi^2/m_n^2)\simeq\Delta m$ is the photon energy in the neutron rest frame. 
In this paper we will use the latest world average of the ``bottle'' measurement of the neutron lifetime to fix its total width, $\tau_n=879.4\pm0.6~\rm s$~\cite{PDG2020}. 
In this case, the branching fraction to $\chi\gamma$ is
\beq
{\rm Br}_{n\ra \chi \gamma} \simeq 0.01\left(\frac{\theta}{5\times10^{-10}}\right)^2\left(\frac{\Delta m}{\rm MeV}\right)^3~.
\label{eq:Brnchigamma}
\eeq

For $m_\chi<m_p+m_e$, the conservation of baryon number ensures that $\chi$ cannot decay. 
For $m_\chi>m_p+m_e$, $\chi$ decays through the weak interaction like the neutron:
\begin{equation}
\begin{aligned}
\Gamma_{\chi\to p e^-\bar \nu}&=\frac{\theta^2}{\tau_n}\left(1-{\rm Br}_{n\ra \chi \gamma}\right)\frac{F(Q_\chi/m_e)}{F(Q_n/m_e)}
\\
&\simeq\frac{1}{9\times10^{22}~\rm s}\left(\frac{\theta}{10^{-10}}\right)^2\frac{F(Q_\chi/m_e)}{F(Q_n/m_e)}~.
\end{aligned}
\label{eq:chipenu}
\end{equation}
Here, $Q_{\chi,n}=m_{\chi,n}-m_p-m_e$ and
\bea
\nn F(x) &=& \frac{\sqrt{x(x+2)}}{60}(2x^4+8x^3+3x^2-10x-15)\\
  &+&\frac{x+1}{4}\ln(1+x+\sqrt{x^2+2x})~
\eea
describes the available phase space. 
Despite not being absolutely stable, $\chi$ can still live longer than the age of the universe $t_U$ for small enough $\theta$, and can therefore potentially make up (part of) the DM even in this part of parameter space. 
For $m_\chi>m_n$, the two-body mode $n\gamma$, induced by Eq.~\eqref{eq:tmm}, opens up. 
The rate for $\chi \to n \gamma$ is of the same form as in Eq.~(\ref{eq:nchigamma}) but with photon energy $\omega=(m_\chi/2)(1-m_n^2/m_\chi^2)$. Numerically,
\bea
 \Gamma_{\chi\ra n \gamma}  &\simeq& \frac{1}{2200~{\rm s}} \bigg(\frac{\theta}{10^{-10}} \bigg)^2 \left|\frac{\Delta m}{10~{\rm MeV}} \right|^3~.
 \label{eq:chingamma}
\eea
Thus, we see that if this channel is open it dominates over the three-body weak decay and that cosmologically interesting $\chi$ lifetimes require smaller $\theta$. We will not discuss the regime $m_\chi>m_n+m_{\pi^0}$ in detail, but note that hadronic modes make up most of the $\chi$ decay width when available and would lead to similar cosmological signatures.

\section{Signals}
\label{sec:signals}
Given the model and rates above, we now turn to the cosmological and astrophysical effects of a dark baryon.
We describe BBN, CMB, and NS signatures in turn.

A crucial ingredient here is the initial $\chi$ abundance, $n_\chi^0=n_\chi(T_0)$, which can depend on the details of cosmic reheating as well as on the strength of the interactions between $\chi$ and the SM. Take, for example, the transition dipole of Eq.~(\ref{eq:tmm}). Through scattering on the SM plasma at a temperature $T$, this leads to a rate of $\chi$-number-change of roughly $\Gamma_{\Delta\chi}\sim \theta^2\mu_n^2 T^3$ which is comparable to the expansion rate of the universe for
\begin{equation}
T\gtrsim 100~{\rm MeV}\left(\frac{10^{-9}}{\theta}\right)^2.
\end{equation}
(Scattering through the strong interaction keeps $\chi$ in chemical equilibrium down to even lower temperatures.) At temperatures above the QCD transition, rather than neutrons one should consider processes involving quarks and $\chi$. As mentioned above, the operator in~(\ref{eq:Lag}) could result from the dimension-6 operator $\bar\chi qqq/\Lambda^2$, giving $\chi$-number-changing rates $\sim T^5/\Lambda^4$, with the cutoff scale $\Lambda^2\propto(\theta\Delta m)^{-1}$. This can keep $\chi$'s in chemical equilibrium down to GeV--PeV temperatures for $\theta\sim 10^{-10}-10^{-20}$ and $\Delta m\sim 1-100~{\rm MeV}$. For $T\gtrsim\Lambda$ further model-dependence is involved since the UV theory that gives rise to the dimension-6 quark-level operator must be used.

So long as the temperature of the universe was at some point above the value at which $\chi$ chemically decouples---which, as we see above, need not be very large---the baryon number abundance would be shared between $\chi$ and SM baryon species. Exact breakdown of total baryon number between $n,p$ and $\chi$ will depend on when the decoupling happens, and the number of dark sector states. (Besides $\chi$ there could be further metastable states with $B=1$.) 
If $\chi$ is the sole ``dark baryon" state, then it is reasonable to take $n_\chi\sim n_p=n_n$ as an initial condition. To display our results, we will focus on two benchmark scenarios: (i) $n_\chi^0=5.4(n_p^0+n_n^0)$ where $\chi$ could make up the entirety of the DM if its lifetime $\tau_\chi\gtrsim t_U$\footnote{In this case $\chi$ decays to nonrelativistic matter and the lower bound of $\sim 20 \ t_U$ on the lifetime of DM decaying to radiation does not simply apply.} which could arise in a more detailed UV completion (see, e.g.,~) and (ii) $n_\chi^0=0.01(n_p^0+n_n^0)$ which could arise if $\chi$ either never came into chemical equilibrium with the SM or dropped out of equilibrium at high temperatures. 

These two benchmarks, for which models with differing high scale physics can be easily incorporated, encapsulate the interesting phenomenological features that we study.

\subsection{BBN}
\label{subsec:boltzmenn}
\begin{figure*}
    \centering
     \includegraphics[width=0.35\textwidth]{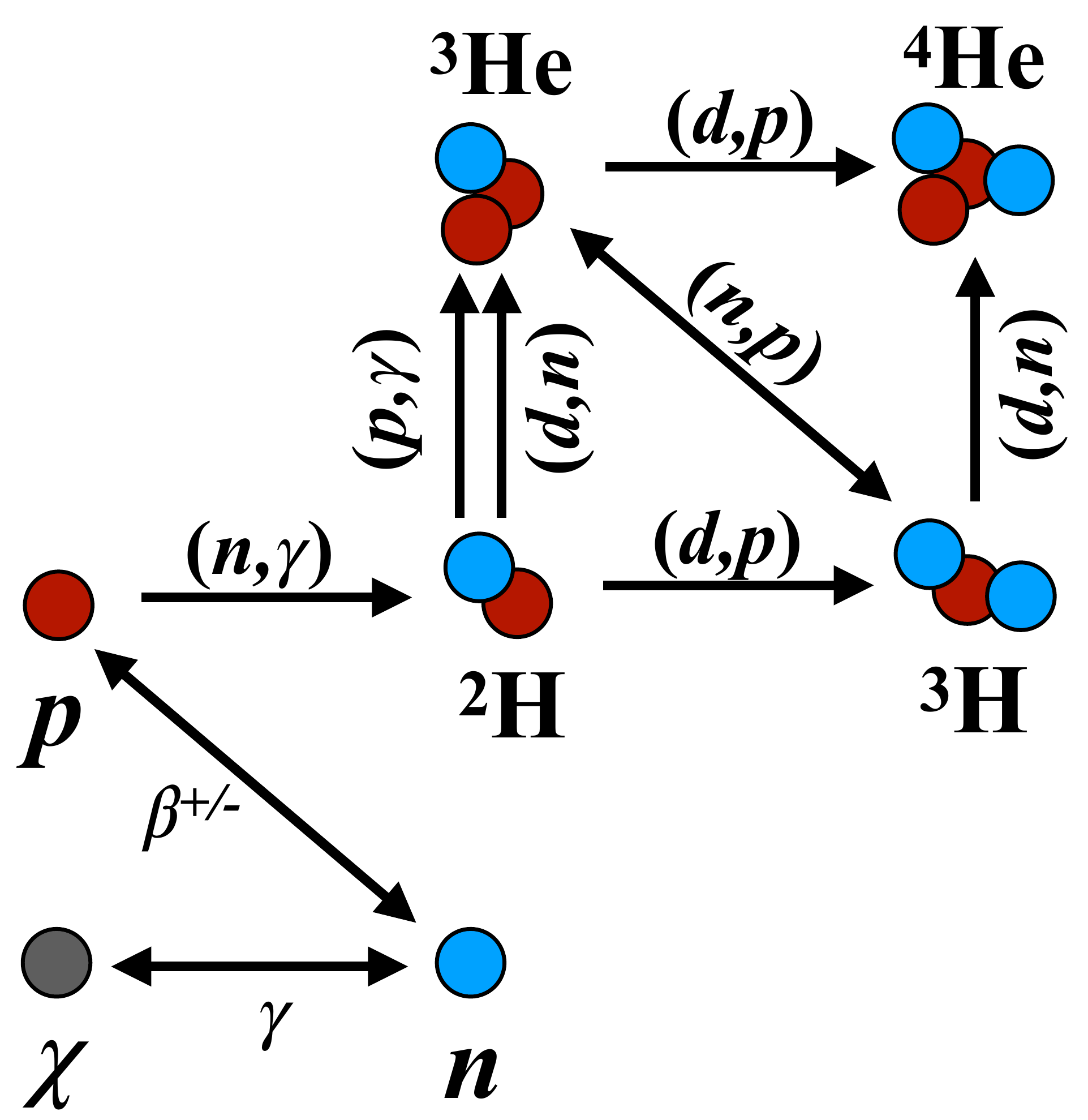} \quad \quad 
     \includegraphics[width=0.46 \textwidth]{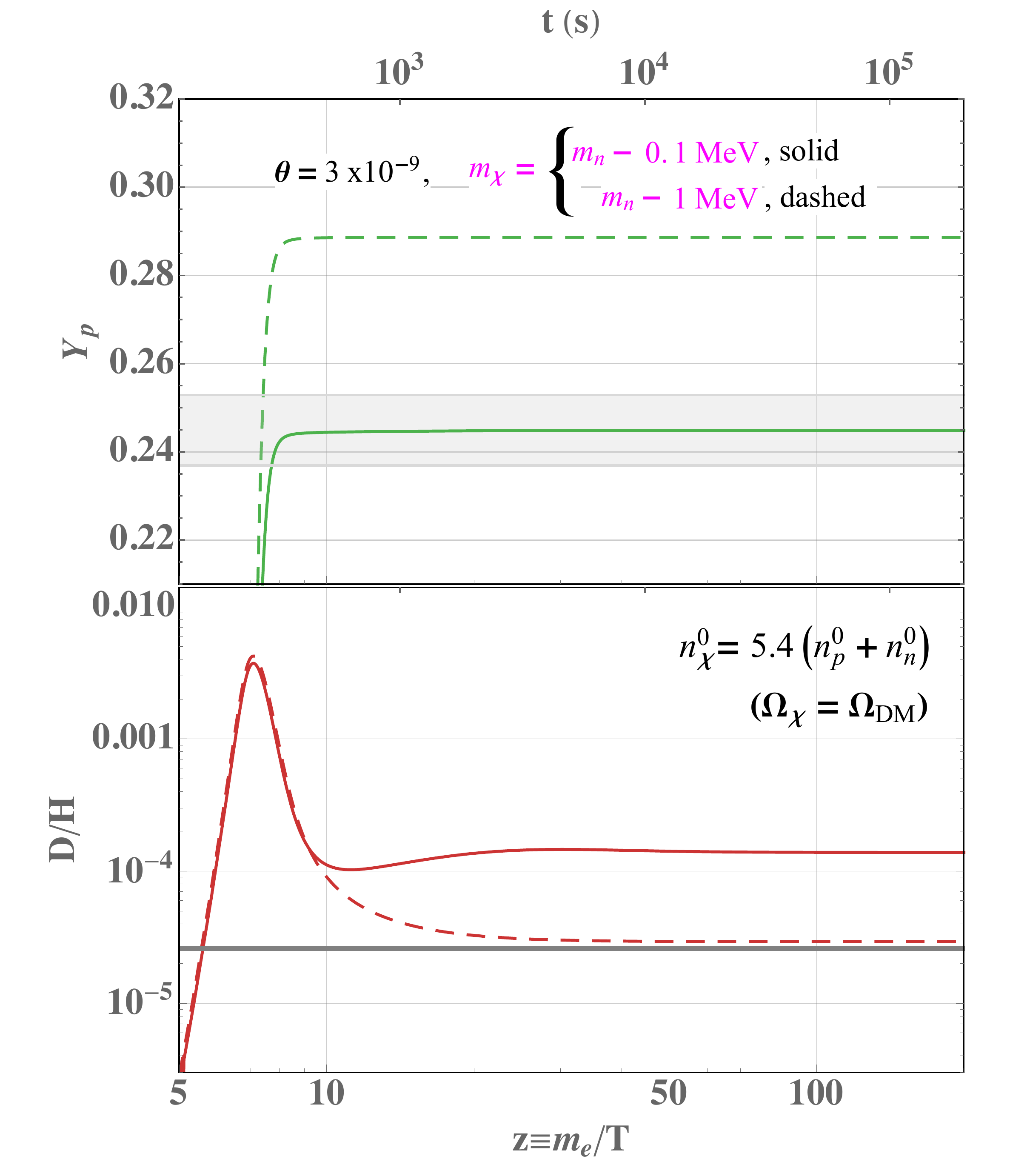} 
        \\
    \caption{ {\bf Left}: The BBN reaction network. 
    Neutron decays to dark baryons $\chi$ could alter the branching fraction of $n$-$p$ conversion beta processes, upsetting the timing of $n$-$p$ freeze-out. 
    Further, $\chi$ can overpopulate neutrons via inverse decays (for $\mdm < m_n$) and direct decays (for $\mdm > m_n$), resulting in overabundances of nuclides.
    {\bf Right}: Helium-4 and deuterium abundances as a function of time or inverse temperature for $\mathbf{\mdm < m_n}$, fixing the $n$-$\chi$ mixing angle $\theta = 3 \times 10^{-9}$, for $\chi$ comprising all of DM. 
    For small $\chi$-$n$ mass splittings inverse decays of $\chi$ greatly overproduce D, while for large splittings the suppression of $n$-$p$ beta processes enhance D and $^4$He abundances.}
    \label{fig:foodweb}
\end{figure*}

The most important parts of the BBN reaction network for 
mass $A=1$--$4$ are shown in the left panel of Fig.~\ref{fig:foodweb}. 
The dark baryon is involved through $n\leftrightarrow\chi$ conversions via the transition dipole of Eq.~(\ref{eq:tmm}). 
Other processes such as $p e^- \to \chi \nu_e$ and $\chi \to p e^- \bar\nu_e$ are extremely slow due to weak- and $\theta$-suppressed rates and hence negligible for BBN.

In the presence of this new interaction, we solve the coupled Boltzmann equations that describe the evolution of the nuclide abundances seen in Fig.~\ref{fig:foodweb}, using fits of thermally averaged weak interaction rates in Ref.~\cite{Serpico:2004gx} and fits of all other thermonuclear reaction rates in Ref.~\cite{Cyburt:2004cq}. 
In addition, we use the temperature evolution in the presence of neutrino decoupling and $e^+e^-$ annihilation from Ref.~\cite{Escudero:2020dfa}. 
With this information we validate our BBN computation by the excellent agreement with measured values of the abundances obtained in the $n_\chi \to 0$ limit.
We describe the main effects for $m_\chi<m_n$ and $ m_\chi>m_n$ below.
\subsubsection{$m_\chi < m_n$}

In this regime the rate for $n\to\chi$ is given by $\Gamma_{n\to\chi \gamma}$ in Eq.~(\ref{eq:nchigamma}) while $\chi\to n$ proceeds through inverse decay,
\beq
 \Gamma_{\chi \gamma \to n} = \frac{\Gamma_{n \to \chi \gamma}}{e^{\Delta m/T}-1}~.
 \label{eq:rate_invDK}
\eeq
In this part of parameter space, $\chi$ is either stable or very long-lived on the scale of BBN. 

To describe the main effects on BBN, it is useful to briefly describe the standard scenario. At high temperatures $n\leftrightarrow p$ reactions keep neutrons and protons in equilibrium. When the temperature drops below $Q_{np}\equiv m_n-m_p=1.29~\rm MeV$, the neutron-to-proton ratio becomes Boltzmann-suppressed with $n_n/n_p\simeq \exp(-Q_{np}/T)$. $n\leftrightarrow p$ reactions become inefficient when their rate, $\Gamma_{np}\propto T^5/\tau_n$, drops below the expansion rate of the universe, $H\propto T^2$. This occurs at $\overline T_{np}\simeq 0.7~\rm MeV$ for $\tau_n=879.4~\rm s$ and the neutron-to-proton ratio freezes out at $\exp(-Q_{np}/\overline T_{np})=0.17$, where, here and in what follows we use bars to denote the values of quantities in the standard scenario. When deuterium begins to be formed, this ratio is slightly reduced by neutron decays so that $\overline{n_n/n_p}\simeq0.14$. Nearly all neutrons available at this time are processed into $^4{\rm He}$ and therefore the resulting $^4{\rm He}$ mass fraction is $\overline Y_p= 2n_n/n_p(1+n_n/n_p)\simeq0.25$.

The presence of the $n \to \chi \gamma$ mode can modify this picture in the following way. 
For fixed neutron lifetime, the $n\leftrightarrow p$ rate, $\Gamma_{np}$, is reduced relative to the standard value by the factor ${\rm Br}_{n\to p} = 1 - {\rm Br}_{n\to\chi}$ since the $n$-$p$ weak coupling must be smaller.\footnote{We mention the impact of other determinations of the weak nucleon coupling in Sec.~\ref{sec:limits}.}
This means that the temperature at which $n\leftrightarrow p$ conversion freezes out is related to the standard value via
\begin{equation}
T_{np}=\frac{\overline T_{np}}{\left(1 - {\rm Br}_{n\to\chi}\right)^{1/3}}\simeq \overline T_{np}\left(1 + \frac{{\rm Br}_{n\to\chi}}{3}\right)~,
\end{equation}
resulting in more neutrons at freeze-out:
\begin{equation}
\begin{aligned}
\frac{n_n}{n_p}&\simeq\exp\left(-\frac{Q_{np}}{T_{np}}\right)\simeq\exp\left[-\frac{Q_{np}}{\overline T_{np}}\left(1-\frac{{\rm Br}_{n\to\chi}}{3}\right)\right]~,
\end{aligned}
\end{equation}
and the fractional increase of this ratio relative to its standard value is therefore
\begin{equation}
\begin{aligned}
\frac{\delta (n_n/n_p)}{\overline{n_n/n_p}}&\simeq \frac{Q_{np}{\rm Br}_{n\to\chi}}{3\overline T_{np}}\simeq 0.5\%\left(\frac{{\rm Br}_{n\to\chi}}{1\%}\right).
\end{aligned}
\end{equation}
Ignoring the small change in the fraction of neutrons that decay before deuterium production, the fractional change in the $^4{\rm He}$ mass fraction is simply
\begin{equation}
\begin{aligned}
\frac{\delta Y_p}{\overline Y_p}&\simeq \frac{\delta (n_n/n_p)}{\overline{n_n/n_p}}\times\frac{1}{1+\overline{n_n/n_p}}
\\
&\simeq 0.4\%\left(\frac{{\rm Br}_{n\to\chi}}{1\%}\right).
\label{eq:fracincYp}
\end{aligned}
\end{equation}
Given the percent-level determination of $Y_p$, this limits the $n\to\chi$ branching ratio to a few percent.

For $\Delta m\lesssim 0.4~\rm MeV$, there is another effect: inverse decays of neutrons through $\chi\gamma\to n$ after the deuterium bottleneck. One should keep in mind that after the bottleneck, the density of free neutrons plummets to low values, and even a small addition of extra neutrons could shift the outcome of reactions for D/H. The left-over neutrons find protons and efficiently produce deuterium that does {\it not} go on to be processed into helium, leading to an extreme increase in the deuterium abundance. This rules out the region of parameter space where ${\rm Br}_{n\to\chi\gamma}=1\%$ for $\Delta m\lesssim 0.4~\rm MeV$. 

The right panel of Fig. 1 shows the $Y_p$ and D/H evolution, as a function of temperature for two representative points of the parameter space. For $n_\chi^0/(n_n^0+n_p^0) = 5.4$ and mixing angle = $3\times 10^{-9}$, a sizeable mass splitting $m_n-m_\chi=1$\,MeV gives D/H close to the standard value, but drastically overproduces $Y_p$. This is a direct result of the modification of weak $n \leftrightarrow p$ conversion at the time of neutron-proton freeze-out. Conversely, a mass splitting of 0.1\,MeV predicts $Y_p$ within its observational range, while giving an overproduction of D/H by a factor $\sim 5$. This is the consequence of inverse decay, $\chi+\gamma\to n$, and incorporation of extra neutrons into D. We conclude that both these parameter points are excluded by the combination of $Y_p$ and D/H.

\subsubsection{$m_\chi > m_n$}

In this mass range, $\chi \to n \gamma$ proceeds directly while $n\to\chi$ goes through inverse decay and is typically Boltzmann-suppressed. For $\tau_\chi\lesssim 0.1~\rm s$, the $\chi$'s decay mostly before the onset of BBN, and there is correspondingly no constraint. For longer lifetimes but with $\tau_\chi\lesssim 10^{13}~{\rm s}$ such that the $\chi$'s decay before the CMB epoch, there is potentially a mismatch in the baryon density $\eta$ as measured in the CMB and through light element abundances. During BBN, some baryon number is sequestered in $\chi$ which mostly plays the role of spectator in this mass range. Each $\chi$ that later decays then produces a baryon which contributes to the baryon density during the time relevant for the CMB. The CMB constraint could be satisfied by dialing down the initial SM baryon and $\chi$ abundance such that their sum is fixed to the CMB value. However, this generally disturbs the agreement between prediction and observation of the deuterium abundance which depends sensitively on the SM baryon density {\it during} BBN. Since the CMB extraction of $\eta$ is at the percent level, this means that the initial $\chi$ density must be smaller than a few percent of the initial SM baryon density.

For a percent-level initial $n_\chi/n_p$, BBN provides nontrivial constraints as a result of the decay $\chi\to n\gamma$. If the decays happen at early times,  this process could change $n\leftrightarrow p$ freeze-out, altering the resulting abundances. The decays after the deuterium bottleneck have two distinct effects: new neutrons are contributed via $\chi$ decay, and non-thermal $\gamma$'s are created, which could split the newly formed nuclei. 
Specifically, when the temperature drops below about $10~\rm keV$,  the electromagnetic cascade induced by the photon produced in $\chi$ decay can lead to the photodissociation of light nuclides, changing the resulting light element abundances in a complicated way.

\begin{figure*}
    \centering
     \includegraphics[width=0.33\textwidth]{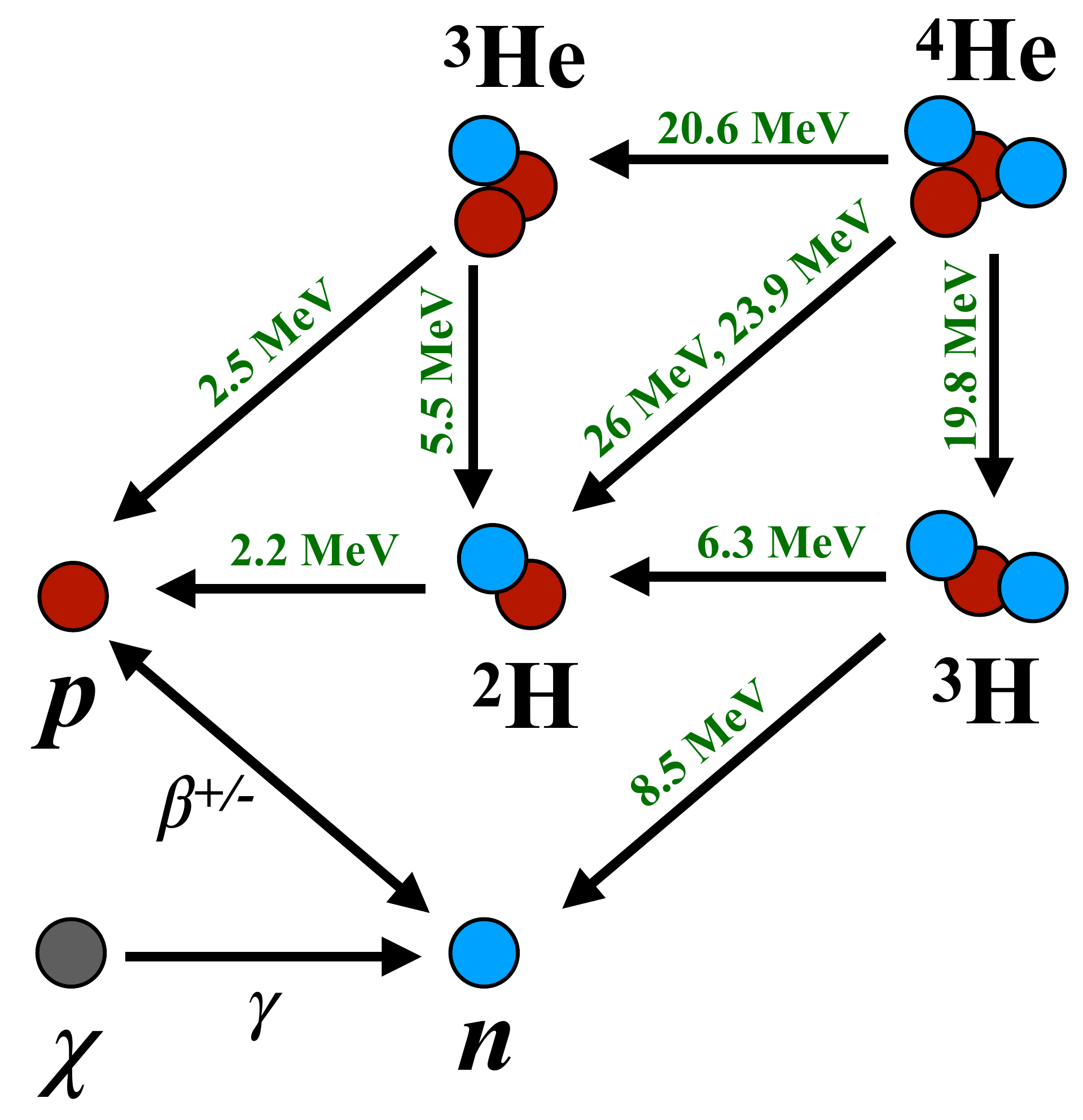} \\ 
          \includegraphics[width=0.46 \textwidth]{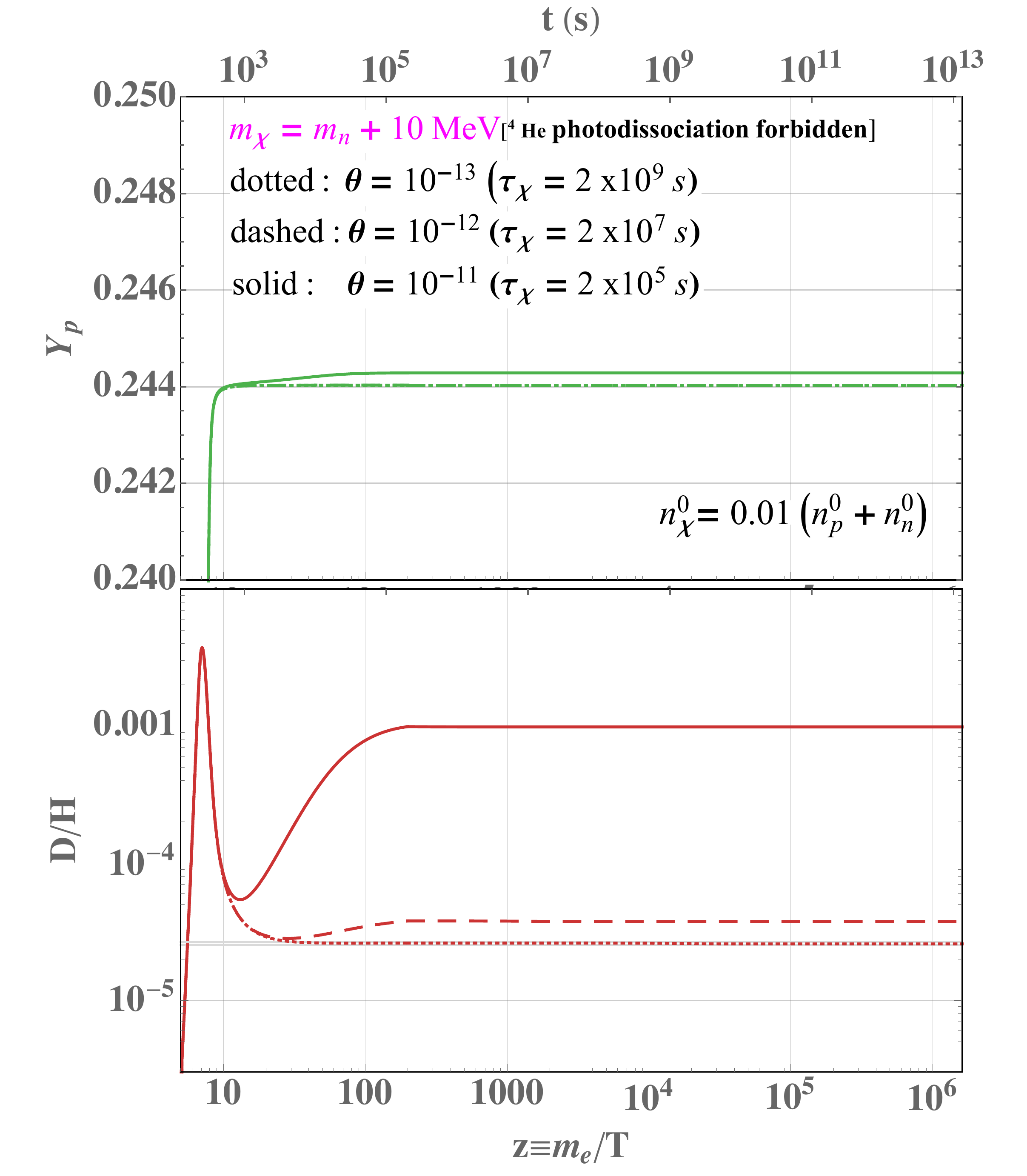} \quad
       \includegraphics[width=0.46 \textwidth]{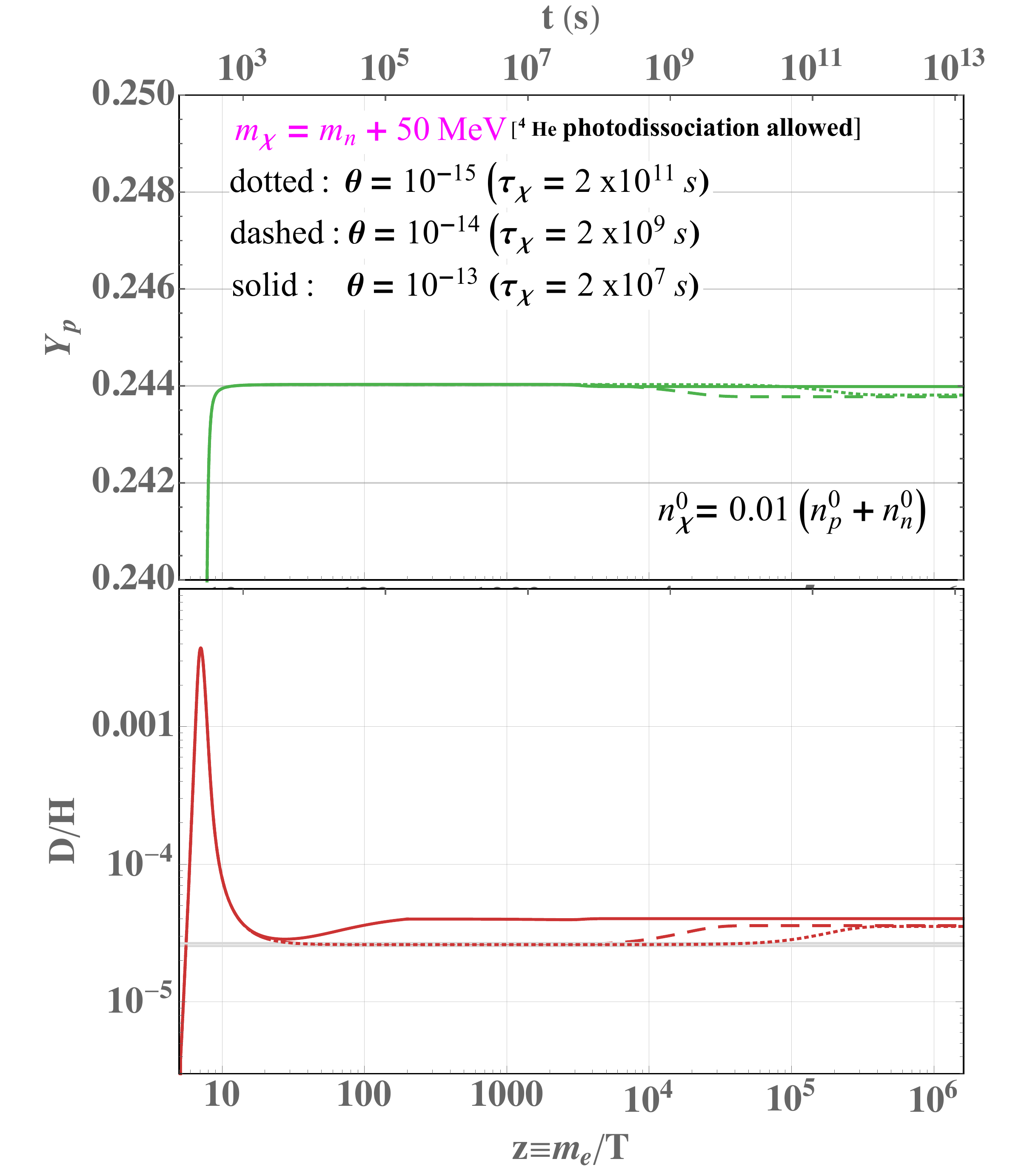} 
    \caption{{\bf Top}: The reaction network for photodissociation of nuclides by the decay $\chi \to n \gamma$ for $\mathbf{m_\chi > m_n}$, relevant for $t > 10^4~$s. 
    The threshold for each photodissociation reaction is labelled.
    {\bf Bottom}: Nuclide abundances as a function of time/inverse temperature for benchmark points corresponding to negligible photodissociation of D [left] and appreciable photodissociation of $^4$He leading to D (and $^3$He) excesses [right].
    An initial $\chi$ abundance of 1\% is assumed, a choice that is safe from CMB measurements of the total baryon density $\Omega_b$. 
    See Sec.~\ref{subsec:boltzmenn} for further details.
    }
    \label{fig:foodwebphotodissoc}
\end{figure*}

The reaction network relevant for this photodissociation stage, indicating the energy threshold for each reaction, is shown in the top panel of Fig.~\ref{fig:foodwebphotodissoc}. To obtain the nuclide abundances we use the photon spectrum $f_\gamma$ derived in Refs.~\cite{Forestell:2018txr,Coffey:2020oir} (which is a more accurate description for sub-100 MeV energy injections than the ``universal spectrum" widely used).
Given a production energy $\omega_{\rm inj} = (m_\chi/2)(1 - m_n^2/m_\chi^2)$, the differential number density of photons per unit energy is then given by
\beq
\mathcal{N_\gamma}(\omega) = \frac{n^0_\chi e^{-t/\tau_\chi}}{\tau_{\chi \to n \gamma}} \bigg[f_\gamma(\omega) + \frac{\delta(\omega - \omega_{\rm inj})}{\Gamma_\gamma(\omega_{\rm inj})} \bigg]~, 
\eeq
where $\Gamma_\gamma$ is the photon relaxation rate taken from Ref.~\cite{Forestell:2018txr}, incorporating $e^+e^-$ production, light-by-light scattering, Compton scattering, and pair-creation on nuclei. Here $n^0_\chi$ is the initial $\chi$ number density. We use the parametric fits obtained in Ref.~\cite{crosssecs_photodissoc} for the photodissociation cross sections.

In the bottom panels of Fig.~\ref{fig:foodwebphotodissoc} we show the evolution of $^4$He and D abundances, fixing $m_\chi - m_n$ = 10 and 50~MeV and varying $\theta$ to scan over $\tau_\chi$.
The initial abundance of $\chi$ is set to 1\% the sum of the initial proton and neutron abundances. For both masses values the final $Y_p$ is well within observed uncertainties.

For $m_\chi - m_n = 10$ MeV, the post-bottleneck stage of BBN at a few$\times 100 - 10^4\,{\rm s}$ is very sensitive to $\chi$ decays to neutrons, which can easily overproduce  D/H. The effect can be appreciable, even if the majority of $\chi\to n\gamma$ decays happen much later. 
As for the non-thermal photon injection, even though $\omega_{\rm inj}\simeq 10~\rm MeV$ is large enough to destroy D, which has a photodissociation threshold of 2.2 MeV, this does not occur efficiently due to an insufficient number of photons produced by the small $\chi$ abundance. 
Photodissociation of $^4$He does not occur either since its threshold is about 20 MeV. Fig. 2 shows that D/H can be sensitive to the combination of lifetimes and abundance of dark baryons at the level of  $\tau_\chi \sim  n_\chi^0/(n_n^0+n_p^0) \times 10^9$\,s. 

For $m_\chi - m_n = 50$ MeV, $\omega_{\rm inj}$ is above the $^4$He photodissociation threshold, and indeed we see a slightly decrease in $Y_p$ when $\chi$ decays occur.
The photodissociation of $^4$He creates D and $^3$He. 
As the $^4$He abundance is $\Oc(10^4)$ greater than those of D and $^3$He, breaking up even a small fraction of $^4$He results in considerable excesses of the latter nuclides.
Thus in the region $m_\chi - m_n \gsim$~20~MeV, our constraints are driven by the excess abundances of D and $^3$He that result from $\chi$ decays. 

For both benchmark masses we observe that larger $\theta$ results in larger D/H due to either larger rates of neutron injection at early times or $^4$He photodissociation at $t \simeq \tau_\chi$.

\subsection{CMB}
\label{subsubsec:cmbsignals}
If dark baryons decay during or after the recombination epoch at around $10^{13}~\rm s$, they can alter CMB observables.  
In both the $m_p+m_e<m_\chi<m_n$ and $m_\chi>m_n$ regimes, the final state in $\chi$ decay contains an energetic electromagnetically interacting particle, respectively either $e^-$ or $\gamma$, that can alter the ionization history of the universe. This has long been recognized as a sensitive probe of particles that decay during or after recombination~\cite{Scott91,*Dodelson:1991bz,*Adams:1998nr,*Chen:2003gz,*Kasuya:2003sm,*Pierpaoli:2003rz}.
Current precision on the CMB temperature anisotropies can access states such as $\chi$ with DM-level energy density for $10^{12}~{\rm s}\lesssim \tau_\chi \lesssim 10^{26}~{\rm s}$.

In addition to distortions of the temperature anisotropies, long-lived particles that source photons before recombination can alter the energy spectrum of CMB photons from that of a simple black body. These photons have different effects on the spectrum depending on the time at which they are injected. For particles decaying to photons after photon-number-changing double Compton scattering becomes inefficient at $\tau_{\rm dC} \simeq 6.1 \times 10^6~{\rm s}$, $\mu$ distortions result. Particle decays after Compton scattering turns off at $\tau_{\rm C} \simeq 8.8 \times 10^9~{\rm s}$ lead to so-called $y =\delta\rho_\gamma/\rho_\gamma$ distortions. These are constrained by the COBE satellite measurement of the CMB spectrum and could be improved by future satellite missions; see, e.g., Ref.~\cite{Chluba:2013wsa}.

\begin{figure*}[t]
\centering
\hspace*{-.5in}
  \includegraphics[width=1\textwidth]{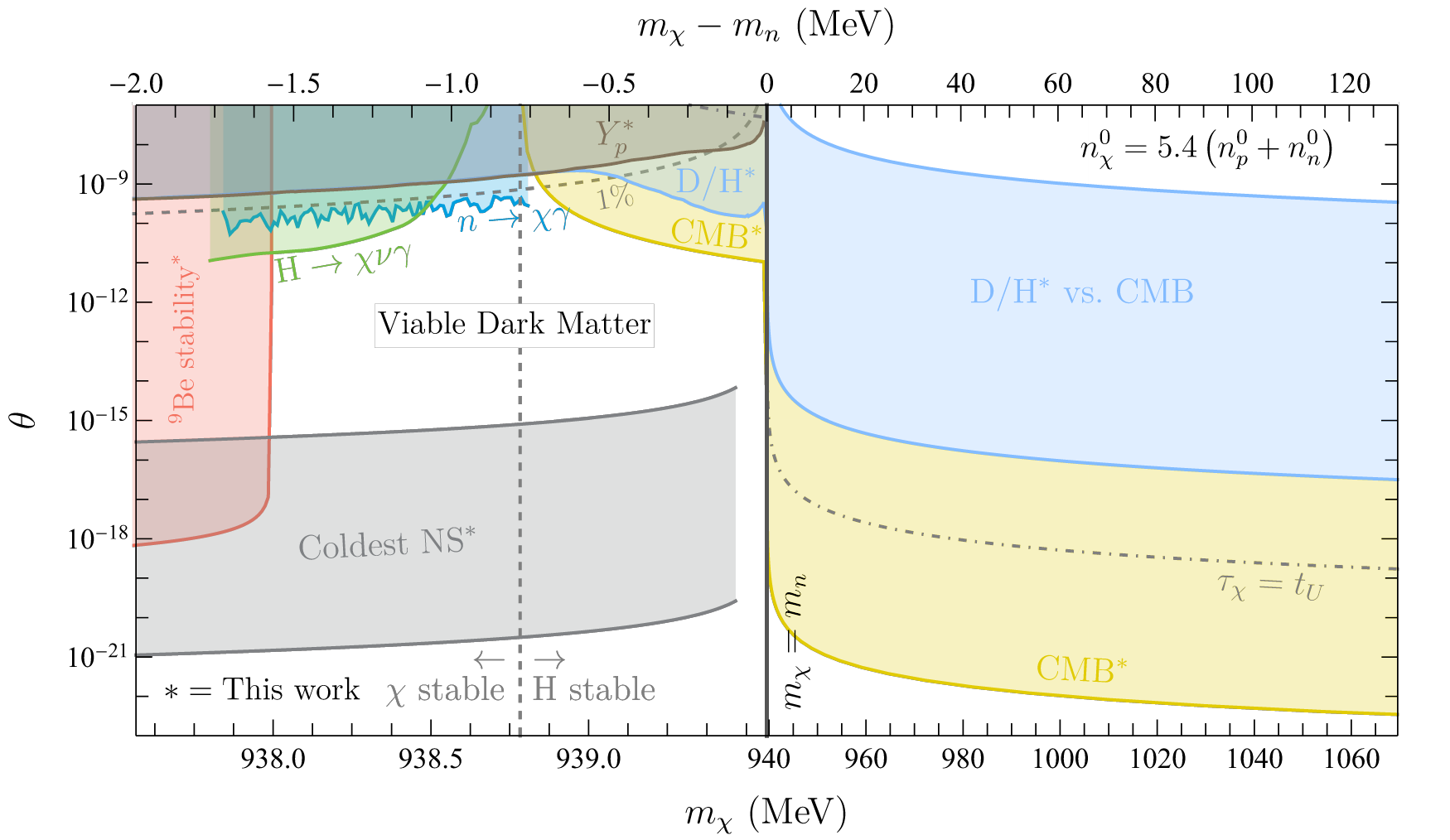}
  \caption{
  Constraints on the $n$-$\chi$ mixing angle $\theta$ as a function of the dark baryon mass $m_\chi$ from various cosmological and astrophysical probes for initial abundance $n_\chi^0 = 5.4\left(n_p^0+n_n^0\right)$, i.e. for $\chi$ making up all dark matter if it is cosmologically long-lived. Note the difference in scale above and below $m_n$. The gray, dash-dotted curves show values of $\theta$ where $\tau_\chi=t_U$, the age of the universe; above (below) this curve the $\chi$ lifetime is shorter (longer). 
  The initial neutron and proton abundances are chosen so that the baryon density at the CMB epoch agrees with observation; this depends on the $\chi$ lifetime, cf. Eq.~(\ref{eq:initialeta}). 
  Asterisks denote limits derived in this paper, which arise from
  (1) the requirement that the $^9{\rm Be}$ lifetime be longer than $t_U$,
  (2) not overheating the coldest neutron star measured, 
  (3) the helium-4 mass fraction, $Y_p$, and the deuterium-to-hydrogen ratio in the region $m_\chi<m_n$, 
  (4) the mismatch between the CMB determination of the baryon-to-photon ratio and the deuterium abundance for $m_\chi> m_n$ given this choice of $n_\chi^0$, $n_p^0$, and $n_n^0$, and
  (5) CMB observations of reionization history from $\chi\to p e^-\bar \nu$ and $\chi\to n\gamma$ decays.
  Existing limits from the UCNA search for the decay $n \to \chi \gamma$ and a recast of Borexino data constraining the radiative decay of atomic hydrogen are also shown. 
  The gray, dashed curve labelled ``1\%" shows where ${\rm Br}_{n\to\chi\gamma}= 1\%$, explaining the neutron lifetime anomaly. 
  See Sec.~\ref{sec:limits} for further details.
           } 
   \label{fig:$}
\end{figure*}

\begin{figure*}[t]
\centering
\hspace*{-.5in}
\includegraphics[width=1\textwidth]{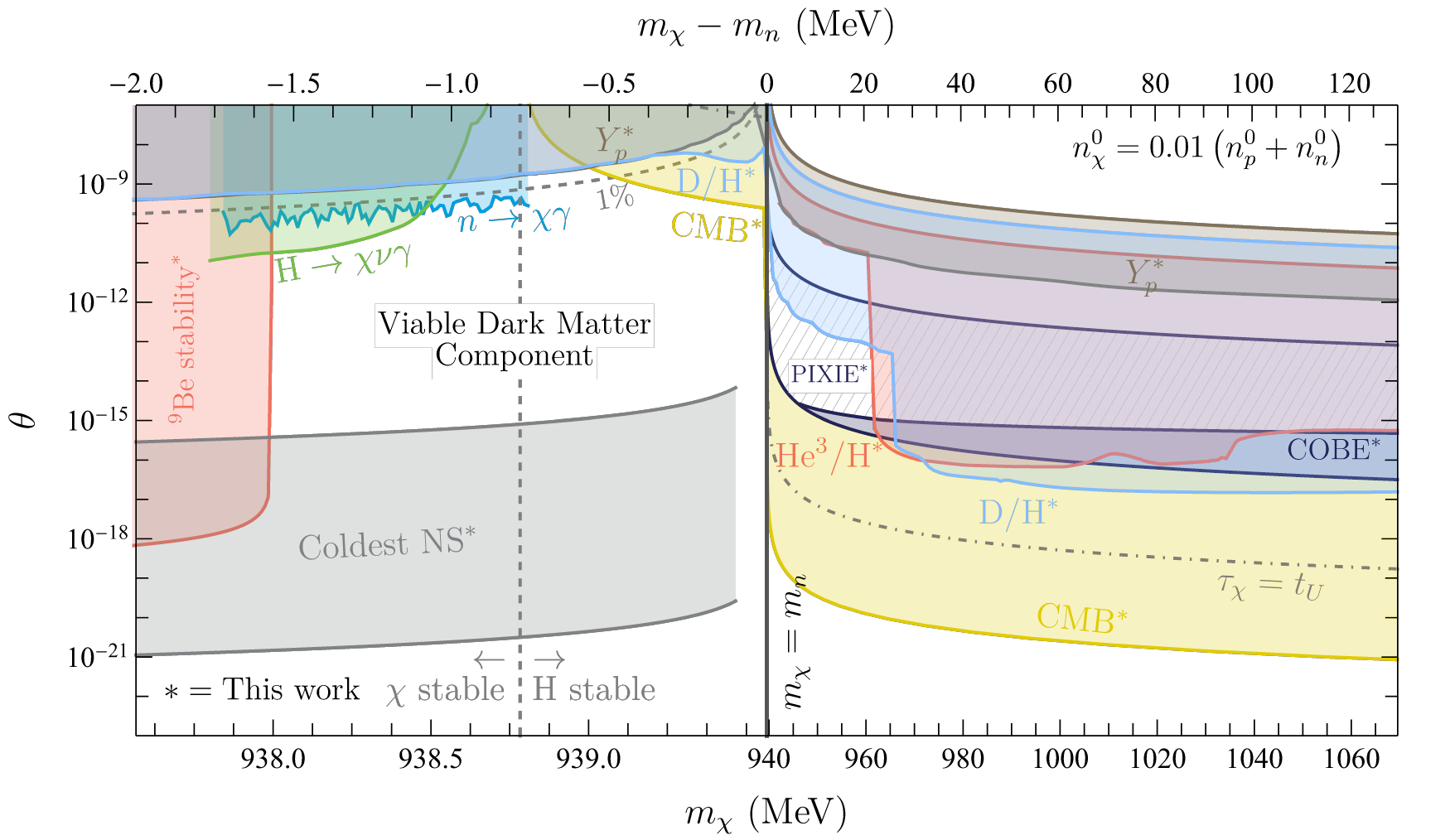}
   \caption{As in Fig.~\ref{fig:$}, but with a reduced initial $\chi$ abundance $n_\chi^0 = 0.01\left(n_p^0+n_n^0\right)$. 
   In this case, for $m_\chi>m_n$ we display limits from primordial abundances of $\rm D$, $^3{\rm He}$, and $^4{\rm He}$, which can come from either photodissociation of $^4{\rm He}$ or enhanced production through $\chi\to n\gamma$. 
   In addition to limits from CMB temperature anisotropy limits, we show those from the COBE measurement of the CMB blackbody frequency spectrum;
   the hatched region could be probed with an improved measurement of this spectrum by the PIXIE satellite.
   See Sec.~\ref{sec:limits} for further details.}
  \label{fig:$$}
\end{figure*}

\subsection{Neutron stars}
\label{subsubsec:NSsignals}

New sensitivity to the parameter space of dark baryons, {\em irrespective of their cosmological abundance}, can be derived from processes in neutron stars. In particular, if $n\to \chi$ conversion is occurring on cosmological time scales, it would lead to a new energy generation mechanism, and could raise the temperatures of NSs. 
This way, the region $10^{-21} \lsim \theta \lsim 10^{-15}$ can be probed by measurements of NS temperatures, and in particular the coldest NS observed, PSR J2144–3933~\cite{coldestNSHST}.
For slow enough decays, the mode $n \to \chi \gamma$ would not populate $\chi$'s within the lifetime of the star, and thus the $\chi$ Fermi sea in the stellar core is largely unfilled, allowing this decay mode to be ongoing. 
These decays would deposit enormous amounts of energy in the star as holes (or vacancies) left behind in the neutron Fermi levels are refilled by the de-excitation of higher-energy neutrons\footnote{Since this mechanism relies principally on removal of neutrons from the Fermi sea, it also applies to models with other exotic neutron decay modes such as $n \to \chi$ + dark photon.}.
This energy deposition raises the stellar luminosity by an amount
\beq
L_{n\to \chi \gamma} = N_n \tilde\Gamma_{n\to\chi \gamma} \Delta E~,
\label{eq:NSlumi}
\eeq
where $\Delta E \approx$ 100 MeV is the average energy deposited per decay obtained from the average neutron Fermi momentum of $\Oc(100)$ MeV,
$N_n = 1.5 \times 10^{57}$ is the number of neutrons in PSR J2144–3933 (whose mass is estimated as 1.4 $M_\odot$),
and $\tilde\Gamma_{n\to\chi \gamma}$ is the decay rate in the dense NS medium accounting for the self-energy $\Sigma$, which gives rise to the effective mass and coupling
\beq
m_n \ra m_n + \Sigma, \ \theta \ra \theta \frac{\Delta m}{\sqrt{\Sigma^2 + \Delta m^2}}~.
\eeq
For our order-of-magnitude estimate, we take an average $\Sigma$ of 10 MeV~\cite{Glendenning:1997wn}.
In the next section we describe constraints arising from this phenomenon.

\section{Constraints}
\label{sec:limits}

Using all the signal estimates above,
we display in Figs.~\ref{fig:$} and \ref{fig:$$} the current constraints and future sensitivities of our setup in the space of $n$-$\chi$ mixing angle $\theta$ versus dark baryon mass $m_\chi$.
The initial $\chi$ abundance is set to $n_\chi^0=5.4(n_p+n_n)$ in Fig.~\ref{fig:$} and $n_\chi^0=0.01(n_p+n_n)$ in Fig.~\ref{fig:$$}. If $\chi$ is cosmologically stable, this corresponds to $\chi$ comprising all of the dark matter or just a percent-level fraction, respectively. For a given ratio of initial densities of $\chi$ to SM baryons, we fix the SM density so that the baryon density at recombination agrees with that measured by the Planck collaboration using CMB data, $\eta_{\rm CMB}=\left(6.13\pm0.04\right)\times10^{-10}$~\cite{Aghanim:2018eyx}. Because $\chi$ decays produce an SM baryon, this means the initial SM baryon density we use depends on $\tau_\chi$. Namely, we set
\begin{equation}
    \frac{n_p^0+n_n^0}{n_\gamma}=\left\{
    \begin{array}{ll}
    \eta_{\rm Planck}, & \tau_\chi>10^{13}~{\rm s}
    \\
    \eta_{\rm Planck}\left(1+\frac{n_\chi^0}{n_p^0+n_n^0}\right)^{-1}, & \tau_\chi<10^{13}~{\rm s}
    \end{array}
    \right..
\label{eq:initialeta}
\end{equation}

For our BBN limits we use the measured values~\cite{Fields_2020}
\bea
\nn Y_p &=& 0.245 \pm 0.004~,\\
{\rm D/H} &=& (2.55 \pm 0.03) \times 10^{-5}~, \\
\nn ^{3}{\rm He/H} &=& (1.0 \pm 0.5) \times 10^{-5}~,
\eea
and quote limits at the 2$\sigma$ level.

We use the results of Refs.~\cite{Slatyer_1211,ClinePoorCMB} for CMB temperature anisotropy limits. For $\tau_\chi$ longer than $t_{\rm rec} \simeq 10^{13}$~s this translates to
\beq
f_\chi/\tau_\chi \lsim 10^{-25}~{\rm s}^{-1}~,
\label{eq:cmbaniso}
\eeq
where $f_\chi$ is the ratio of the $\chi$ energy density to that measured in DM.
For shorter lifetimes the bound weakens rapidly, disappearing for $\tau_\chi\lesssim 10^7~\rm s$.
The above constraint is roughly constant for energy injections $1~{\rm keV}\lsim\omega_{\rm inj}\lsim 100~{\rm MeV}$ and comes from studies of $e^+e^-$ or $\gamma$ injection via DM decay. For $m_\chi>m_n$, $\chi$ decay produces a photon and this limit applies directly. For $m_p + m_e < m_\chi < m_n$, the energetic, interacting particle in the final state is a single electron; however, we adopt the above limit as an order-of-magnitude estimate.
Note that for energy depositions of 200--250 MeV, Milky Way gamma-ray background measurements give lifetime limits quite similar to the CMB limit above~\cite{Cohen:2016uyg}, and would apply to larger $m_\chi$.

To obtain exclusion limits on $\theta$ from NS heating we use Eq.~\eqref{eq:NSlumi} and demand 
\beq
L_{\rm n\to \chi \gamma} \leq 4\pi \sigma_{\rm SB} R^2_\star T^4_\star~, 
\eeq
where $R_\star = 11$ km and $T_\star = 4.3 \times 10^4$~K are respectively the radius estimate and upper bound on blackbody temperature of PSR J2144–3933. The NS heating constraints are shown by a horizontal grey band on Fig. 3 and 4 that covers the $\theta \sim 10^{-20}-10^{-16}$ range for $m_n>m_\chi$.
The ceiling on $\theta$ is obtained by simply demanding that the decays are longer than the age of the star, estimated to be $3 \times 10^8$~yr.
This ceiling  could be improved by luminosity measurements of younger NSs; similarly, the floor on $\theta$ could be improved by future measurements of old NSs' temperatures down to $\Oc(10^2-10^3)$~K in infrared telescopes~\cite{Baryakhtar:2017dbj,*Raj:2017wrv,*Garani:2018kkd,*Acevedo:2019agu,*Bell:2018pkk,*Camargo:2019wou,*Hamaguchi:2019oev,*Bell:2019pyc,*Garani:2019fpa,*Joglekar:2019vzy,*Keung:2020teb,*Joglekar:2020liw,*Bell:2020jou,*Garani:2020wge}; for example, observing a 1000~K NS would imply a $\chi$ lifetime bound $(42000/1000)^4 \sim 10^6$ times stronger, i.e. a $\theta$ bound $\sim 10^3$ times stronger.
In forthcoming work~\cite{NS-MPR} we undertake these tasks, as well as a more careful treatment of density effects, equation-of-state (EoS) uncertainties, and other effects. 
We would like to emphasize that unlike the case with accumulating/annihilating DM~\cite{Baryakhtar:2017dbj,*Raj:2017wrv,*Garani:2018kkd,*Acevedo:2019agu,*Bell:2018pkk,*Camargo:2019wou,*Hamaguchi:2019oev,*Bell:2019pyc,*Garani:2019fpa,*Joglekar:2019vzy,*Keung:2020teb,*Joglekar:2020liw,*Bell:2020jou,*Garani:2020wge}, where actual sensitivity to DM is predicated on future observational progress, dark baryons are {\em already} constrained by existing NS temperature measurements, and regardless of whether they form DM. 

We also display the constraint from the stability of $^9{\rm Be}$ due to neutron decays inside weakly bound $^9$Be. As pointed out in~Ref.~\cite{ExoticNucleiDecay}, if $m_\chi<m_{^9{\rm Be}}-2m_\alpha=938.0~\rm MeV$, with $\alpha$ the $^4{\rm He}$ nucleus, the decay $^9{\rm Be}\to\chi\alpha\alpha$ can proceed via
\begin{equation}
^9{\rm Be} \to 2\,^4{\rm He} +\chi;~~^9{\rm Be} \to 2\,^4{\rm He} +\chi + \gamma
\label{Be9}
\end{equation} 
reactions. The energy release in both cases is
given by 
\begin{equation}
    Q_{^9{\rm Be}} = m_{^9{\rm Be}}-2m_\alpha -m_\chi =\Delta m -  1.574\,{\rm MeV}.
\end{equation}
 Both reaction rates are proportional to the cube of the momentum of lightest emitted particle, which translates to a different scaling with $Q_{^9{\rm Be}}$: 3/2-power for the first rate, and 1/3 for the second rate. This makes the first rate vastly larger, which we estimate  using a simple $^8{\rm Be}$-$n$ cluster model for $^9{\rm Be}$ where the neutron has binding energy $E_b$. Taking the perturbation that mediates the transition as the mixing angle times the potential energy of the $^8{\rm Be}$-$n$ system, we reduce the resulting transition rate,
 \begin{equation}
    \Gamma_{^9{\rm Be}} \simeq \frac{\theta^2 m_\chi p_\chi(E_\chi+|E_b|)^2}{4\pi^2}
    \times |\langle n|\chi \rangle |^2 d\Omega_{\vec{p_\chi}},
\end{equation}
to a simple projection of bound $|n\rangle $ state onto the final state wave function $|\chi\rangle$. The energy of the final state dark baryon $E_\chi$ is approximated by $Q_{^9{\rm Be}}$. Taking the neutron wave function in the $p_{3/2}$ wave, and ensuring that its spatial extent is similar to the $^{9}{\rm Be}$ nuclear radius, in the limit of small energy release we estimate the decay rate as
 \begin{equation}
 \Gamma_{^9{\rm Be}} \sim 50\,{\rm keV} \times \theta^2 \times \left(\frac{Q_{^9{\rm Be}}}{1\,{\rm MeV}}\right)^{3/2}~.
 \label{Be9decay}
 \end{equation}
 Consequently our prediction for the lifetime of $^9{\rm Be}$ is
 \begin{equation}
     \tau_{^9{\rm Be}} \sim 4\times10^{10}~{\rm yr}\left(\frac{10^{-19}}{\theta}\right)^2\left(\frac{1\,\rm MeV}{Q_{^9{\rm Be}}}\right)^{3/2}~,
 \end{equation}
to be compared with the age of extremely metal-poor (thus very old) stars in which the beryllium-9 atoms are observed~\cite{Rich:2009gj}. Very conservatively requiring $ \tau_{^9{\rm Be}} <3\times 10^9\,{\rm yr}$, which would give $\sim O(100)$ suppression of beryllium abundance in the oldest stars, we arrive at the stability limits shown in Fig.\,3 and 4. Notice the significant overlap of NS and beryllium lifetime constraints, which provides an upper limit on the $\Delta m$ parameter for a very wide range of $\theta$.
Also, fortuitously, the combined NS and $^9{\rm Be}$ lifetime exclusion region does not depend on an order-of-magnitude variation in the limiting value for $ \tau_{^9{\rm Be}} $ and/or in the nuclear decay rate (\ref{Be9decay}). 
We note in passing that superior limits on beryllium lifetime can be derived in the laboratory by exploiting the ionization/heat created by outgoing $\alpha$-particles. 

We also show the regions where $\chi$ (against decay to $p e^- \bar\nu_e$) and
H (against decay to $\chi \nu_e$) are stable. 
For reference we have shown the contour along which the branching fraction of $n \ra \chi \gamma$ is 1\%, as required to explain the neutron lifetime anomaly.

In the $m_\chi<m_n$ regime for both initial $\chi$ densities displayed, the BBN constraints largely follow a contour of constant $n\to\chi\gamma$ rate, through its effect on $n\leftrightarrow p$ freeze out (for fixed $\tau_n$). As mentioned previously, for mass splittings less than a few hundred keV, production of neutrons after the deuterium bottleneck through inverse decay is important and strengthens the limit. In this regime, the limits from CMB anisotropies on EM energy injections during and after recombination are quite important for splittings less than several hundred keV. 

For comparison we also show limits in this mass range that come from a search for $n \to \chi \gamma$ using ultracold neutrons at LANL~\cite{ucnGammaLANL} and a recast~\cite{HDK} of Borexino data~\cite{Borexino} to constrain the radiative decay of atomic hydrogen to $\chi \bar\nu_e \gamma$. We also show in both figures the value of the mixing angle that would give rise to ${\rm Br}_{n\to\chi\gamma}=1\%$, which would explain the neutron lifetime anomaly. While terrestrial data rules out large parts of this parameter space for mass splittings above roughly $700~\rm keV$, cosmological data from the CMB and BBN probe larger $m_\chi$, even for a small initial $\chi$ abundance, where the small $\Delta m$ leaves little energy for visible particles produced in association with $\chi$ in neutron decay.

For $m_\chi>m_n$, $\chi\to n\gamma$ opens up and, as a result, CMB anisotropy limits apply to small mixing angles, for $10^{-22}\lesssim\theta\lesssim10^{-16}$. For larger mixing angles in this mass regime, $\chi$'s decay before the CMB epoch and therefore contribute to the baryon density measured in the CMB. As seen in Eq.~(\ref{eq:initialeta}), this means we have to set $(n_\chi^0+n_p^0+n_n^0)/n_\gamma^0=\eta_{\rm CMB}$. For large $n_\chi^0/(n_p^0+n_n^0)$, if $\chi$ lives long enough such that it decays after BBN, this means that the SM baryon density during BBN is much smaller than in the standard case and disturbs the good agreement with observational data, particularly the primordial deuterium abundance. This is seen to be the case in Fig.~\ref{fig:$$} where $n_\chi^0/(n_p^0+n_n^0)=5.4$.

In the case of a smaller initial $\chi$ abundance, the $m_\chi>m_n$ regime can have more interesting consequences. CMB temperature anisotropies still constrain small mixing angles where $\tau_\chi\gtrsim10^{12}~\rm s$. Shorter lifetimes can impact the apparent blackbody spectrum of CMB photons. We show constraints from the COBE satellite as well as the region that could be probed by the proposed PIXIE experiment, which mostly come from the so-called $y$ distortions that occur when $\chi$ decays after Compton scattering freezes out at $\tau_C$~\cite{Chluba:2018cww,Forestell:2018txr}. In addition, there are important BBN limits. For relatively long lifetimes up to $10^{13}~\rm s$, late $\chi\to n\gamma$ decays photo-dissociate light elements, and the resulting limits are driven largely by not disturbing the agreement with expectation of the measured  D/H ratio; these limits tend to follow contours of constant $\tau_\chi$, up to effects due to thresholds. For shorter lifetimes, down to $\tau_\chi\gtrsim 4 s$, the injection of neutrons at late times can change the synthesis of light elements; the maximum $\theta$ probed here is determined by the $^4{\rm He}$ abundance.

We now remark on some limits that are not displayed here. 
Measurements of the diffuse gamma-ray background constrain $\chi$ decays, albeit more weakly than CMB limits we have shown on the reionization history of the universe~\cite{BerezhianinHDK}. 
Additionally, the impact on the CMB from just $\chi$ decay, ignoring its decay products, is much less important because $\chi$ decays to nonrelativistic matter and electromagnetically interacting particles.

Finally, Ref.~\cite{Marciano_gA} derived an upper limit on the neutron's branching fraction to exotic modes of 0.27\% at 95\% C.L. This limit was obtained by using the most up-to-date radiative corrections as well as measurements of the nucleon axial coupling $g_A$ from nuclear $\beta$-decay asymmetries, comparing the predicted neutron $\beta$-decay lifetime to the world average of the ``bottle'' measurements. This limit would appear very close to the ${\rm Br}_{n\to\chi\gamma}=1\%$ contour in Fig.~\ref{fig:$} or \ref{fig:$$}. As emphasized in Ref.~\cite{BerezhianinHDK}, any solution to the neutron lifetime anomaly using exotic neutron decay would then imply a tension with recent measurements of $g_A$, which would be further highlighted if evidence were found for the dark baryon in this part of parameter space.

\section{Discussion}
\label{sec:concs}

Our results in the $\mdm < m_n$ region for $\Omega_\chi = \Omega_{\rm DM}$ in Fig.~\ref{fig:$} show that only a narrow 100 keV window between $\mdm \simeq$ 938.8$-$938.9~MeV survives that could explain the neutron lifetime anomaly, motivating laboratory searches to increase their sensitivity to cover this window and settle the question.
This window is only 300 keV-wide for $n_\chi^0=0.01(n_p+n_n)$ in Fig.~\ref{fig:$$}, implying that experimental improvements will probe this scenario even for percent-level DM populations of dark baryons. 
We remark here that if an additional decay mode to a dark photon $n \to \chi A_{\rm D}$ were to exist, as present when large self-interactions of $\chi$ mediated by $A_{\rm D}$ are required to evade NS EoS constraints~\cite{ClineCornellCosmo}, then a 1\% total exotic branching fraction would correspond to smaller $\theta$ than shown in our plots, widening the surviving window.

Analogously, for $n_\chi^0=0.01(n_p+n_n)$ only a small window is open in the range $0 < \mdm - m_n < 20$~MeV, which will be probed by the PIXIE experiment in the CMB frequency spectrum.
Other parts of parameter space may be probed in the future via exotic decays of promising nuclei such as $^{11}$Be~\cite{ExoticNucleiDecay}, 
nuclear capture of $\chi$ at dark matter and neutrino experiments~\cite{UtahNuclearCapture},
and measurements of neutron star luminosities at $\Oc(10^2-10^3)$K temperatures~\cite{Baryakhtar:2017dbj,*Raj:2017wrv,*Garani:2018kkd,*Acevedo:2019agu,*Bell:2018pkk,*Camargo:2019wou,*Hamaguchi:2019oev,*Bell:2019pyc,*Garani:2019fpa,*Joglekar:2019vzy,*Keung:2020teb,*Joglekar:2020liw,*Bell:2020jou,*Garani:2020wge}.
In non-minimal setups, the ``hydrogen portal" could be relevant: future DM experiments could detect the decays of H in the Earth's hydrosphere and the Sun~\cite{X1TexcessHportal}.

Our setup in the $\mdm > m_n$ regime could potentially address the lithium problem, the long-standing discrepancy in the $^7$Li abundance between standard BBN predictions and low-metallicity stars' Spite plateau measurements~\cite{Fields_2020}:
\beq
^7{\rm Li}/{\rm H} = \begin{cases} 
4.7 \pm 0.7 \times 10^{-10}, \  \ \ {\rm standard \ BBN}~,\\
(1.6 \pm 0.3) \times 10^{-10}, \ {\rm Spite \ plateau}~.
\end{cases}
\eeq
These values account for the fact that $^7$Be created during BBN is later converted to $^7$Li via electron capture. 
As reviewed in Ref.~\cite{ReviewBBNPospelovPradler}, an injection of $\gsim 10^{-5}$ neutrons per baryon from $\chi \to n \gamma$ just after $^7$Be synthesis via $^3{\rm He} + ^4{\rm He} \to ^7{\rm Be} + \gamma$ (i.e. $3\times 10^2~{\rm s} < t < 10^3~{\rm s}$ or 60 keV$> T >$ 30 keV) would burn $^7$Be via the reaction $^7{\rm Be} + n \to ^7{\rm Li} + p$ and bring the $^7$Be+$^7$Li abundance down to the observed values.
For our $n^0_{\chi} = 0.01 (n^0_p + n^0_n)$ scenario (that is not ruled out by CMB $\eta_b$ measurements), we obtain $n_n/n_p > 10^{-5}$ injection within $t < 10^3$~s if the $\chi$ lifetime $\tau_\chi \lsim 10^6$~s.
Unfortunately such short lifetimes are deeply excluded by excess D/H in our setup, as in many other solutions that utilize ``extra neutrons"~\cite{Coc:2014gia}. 
For example, in Fig.~\ref{fig:$$}, $\tau_\chi = 10^6$~s corresponds to a contour just above the PIXIE reach, which is in the excluded region.
One could naively imagine solving the lithium problem while preserving D/H by photodissociating excess deuterium down to observed levels, achieved by increasing $n^0_{\chi}/(n^0_p+n^0_n)$ and consequently increasing $\gamma$ injection via $\chi \to n \gamma$. (Other solutions to lithium problem with unstable particles, where neutron-injection-induced increase in D/H is followed by its reduction via the electromagnetic injection, are known \cite{Pospelov:2010cw}.)
But as we had discussed in detail, increasing $n^0_{\chi}/(n^0_p+n^0_n)$ above the percent level for the dark baryon model at $\tau_\chi < t_{\rm rec}$ would run afoul of the agreement between the CMB $\Omega_b$ measurement with the observed D/H ratio. Thus we conclude that the minimal dark baryon model cannot reduce primordial $^7{\rm Li}$ abundance. 

We finally note that a relatively large portion of parameter space exists that might be compatible with the dark baryon $\chi$ being the entirety of DM; see Fig.~3. 
Since it is one of the relatively economical/natural models, this topic would warrant future investigations~\cite{NS-MPR}. 
In particular, this entire area induces a relatively fast $n\leftrightarrow \chi$ equilibration inside neutron stars. 
And while the NS mass-radius relation in equilibrium can be brought to within observed ranges using a repulsive dark force for $\chi$, it is far from certain how quickly this equilibrium is achieved, and if there are any other outstanding observational consequences of $n\leftrightarrow \chi$ equilibration processes that could disfavor this part of the parameter space.

\section*{Acknowledgments}

We are indebted to David Morrissey for providing us with photon spectra from electromagnetic energy injection, and for fruitful conversations.
The work of D.\,M. and N.\,R. is supported by the Natural Sciences and Engineering Research Council of Canada. 
T\acro{RIUMF} receives federal funding via a contribution agreement with the National Research Council Canada.
M.P. is supported in part by U.S. Department of Energy Grant No.
desc0011842.

\bibliography{refs}

\end{document}